\documentclass{article}
\usepackage{microtype}

\DisableLigatures{encoding = *, family = * }
\usepackage{amsmath}
\usepackage{algorithm}
\usepackage{algpseudocode}
\usepackage{algpascal}
\usepackage[hidelinks]{hyperref}
\usepackage{algorithm}
\usepackage{algpseudocode}
\usepackage{algpascal}
\usepackage{subfigure}%
\usepackage{float}
\usepackage{makeidx}
\usepackage{balance}
\usepackage{graphicx}
\usepackage{lscape}
\usepackage{multirow}
\usepackage{multicol}
\usepackage{breqn}
\usepackage{verbatim}
\usepackage{subfigure}
\usepackage{newclude}
\usepackage{array}
\usepackage{booktabs}
\usepackage{graphicx}
\usepackage{caption}
\usepackage{acronym}
\usepackage{footnote}
\usepackage{subfigure}
\usepackage{eurosym}
\usepackage{float,lscape}

\acrodef{YoMo}{You Only Measure Once}

\usepackage{pgfplots}
\pgfplotsset{compat=newest}
\usetikzlibrary{plotmarks}

\usepackage{acronym}

\acrodef{GPL}{Gnu Public License}

\usepackage{hyperref}
\hypersetup{
  pdfauthor={{Christoph Klemenjak, Dominik Egarter, Wilfried Elmenreich}},
  pdfkeywords={Power meter, smart meter, smart plug, open-source, open-hardware, low-cost},
  pdftitle={YoMo - The Arduino-based Smart Metering Board}
}

\begin{document}

\title{YoMo - The Arduino-based Smart Metering Board}
 \author{
Christoph~Klemenjak,~Dominik~Egarter, and~Wilfried~Elmenreich\\
Institute of Networked and Embedded Systems\\
 Alpen-Adria-Universit\"at Klagenfurt, Austria\\
 \{\emph{name.surname}\}@aau.at
 }

\date{}    

\maketitle

\begin{abstract}
Smart meters are an enabling technology for many smart grid applications. This paper introduces a design for a low-cost smart meter system as well as the fundamentals of smart metering. The smart meter platform, provided as open hardware, is designed with a connector interface compatible to the {Arduino} platform, thus opening the possibilities for smart meters with flexible hardware and computation features, starting from low-cost 8 bit micro controllers up to powerful single board computers that can run Linux. The metering platform features a current transformer which allows a non-intrusive installation of the current measurement unit.
The suggested design can switch loads, offers a variable sampling frequency, and provides measurement data such as active power, reactive and apparent power. Results indicate that measurement accuracy and resolution of the proposed metering platform are sufficient for a range of different applications and loads from a few watts up to five kilowatts.
\end{abstract}

\section{Introduction}

A smart meter is an electronic device that records consumption of energy in regular intervals and makes that information available to its stakeholders by a network interface. A smart meter enables the possibility to provide immediate feedback on one's energy consumption, greatly reduces the effort of reading the meter value, increases the grid state awareness, supports the introduction of time-based energy tariffs, and possibly supports a quick access for switching on/off the energy flow. Besides as a device used for energy billing by utility companies, smart meters can also become a central component in future, autonomous energy systems, e.g., a smart microgrid~\cite{sobe:informatik12} with local energy production that needs to be monitored and coordinated.

In order to support these applications, there is a need for a metering unit that is able to monitor measurement quantities such as voltage, current, active and reactive power. The power consumption of the metering device itself should be low in order to save energy consumption, provide reasonable run-time with batteries and to restrict measurement interference the device is powered by the same grid that is measured. The device should provide an adjustable sampling frequency in order to support long-time measurements (with a low sample frequency in order to reduce the number of measurements to be stored) as well as high sample frequencies for device detection mechanisms based on signal processing algorithms~\cite{Egarter2014}.
Many applications such as remote or timed control of devices or load shedding require also the possibility of switching the connected load on or off.
Such a smart meter will be equipped with a local processing unit for processing and storing measured data and a communication unit allowing remote access to the measurement data and an access to configuration and maintenance interfaces~\cite{TUW-136763}.
While commercial smart meters are typically closed systems with a well-defined but limited range of applications, we aim at designing an open and low-cost metering device which is adaptable for different purposes. The device firmware and application software will be published as open source (GNU-license - {GPLv3}), while the hardware design will be made available under a free documentation license (Common Creative license {CC-BY-4.0}\footnote{http://creativecommons.org/licenses/by/4.0/}).

In this paper, we present the design of a smart meter which fulfils the requirements stated above. Our smart meter\footnote{http://yomo.sourceforge.net}, \ac{YoMo}, is provided as a shield, i.e., an extension, for an Arduino microcontroller board. The metering shield contains a measurement unit, circuitry for galvanic isolation from the measured circuit, signal conditioning and measurement hardware. The designed PCB\footnote{PCB - printed circuit board} can be attached and used together with different Arduino footprint-compatible boards.
The provided smart meter designed should not be a replacement for existing smart meters installed in homes. It should be a possible development platform for industry and academics to introduce, to develop and to test future smart metering applications with the possibility to measure several household's power demand parameters on different power levels starting on appliance level demand going up to the household demand.
 In a comparison of existing open-source metering solutions we show that our proposed design has a unique set of useful features. After introducing the rationale behind the \ac{YoMo} design, we present its application in a low-cost energy monitoring system featuring distributed measurement architecture featuring a \ac{YoMo}/Arduino component together with a more performant Raspberry Pi embedded computer which is used to store and to present the measurement in a graphical user interface provided by a small webserver. An evaluation of the measurement accuracy and experiences with the \ac{YoMo} board in the energy monitoring application have shown promising results for the applicability of our low-cost smart meter.

\section{Related Work} \label{sec:related}
As smart metering is one of the key points for the future smart grid and the way to improve energy awareness of homes, several commercial products monitoring the homes energy consumption exist.
Examples are {Wattson}\footnote{http://wattson.ch/}, {CurrentCost}\footnote{http://www.currentcost.com/} or {TED}\footnote{http://www.theenergydetective.com/} power meters.
These solutions provide measurements of the household power demand whereas {Plugwise}\footnote{http://www.plugwise.com/} or {KillAWatt}\footnote{http://www.p3international.com/products/p4400.html} are monitoring on appliance or socket level.
These commercial products are typically very limited in its applicability,
lacking in sense of variability of the measurement frequency, adaptability and programmability.

Additionally, there exist a few, non-commercial open-source metering solutions such as the open-energy-monitor\footnote{http://openenergymonitor.org/emon/}.
This monitoring solution offers an open-source metering solution, where the monitoring shield is operated either with an Arduino board or individually.
The metering unit communicates its measurement to a central station which can be for example a Raspberry Pi.
On the central station a visualization tool can be installed to show the measurement in real-time.
Unfortunately, the OEM only provides active and apparent power, uses non-common smart home communication technology and cannot switch devices on or off.
Recent research concentrates on building an open-source smart meter and power monitoring unit.
Popular examples therefore are {ACMe} \cite{Jiang2009} from Berkeley and {Plug} from the MIT \cite{Lifton2007}.
More recent metering solutions are the cheap smart meter ({CSM}) \cite{ellerbrock:2012}, the {SmartMETER.Kom} \cite{reinhardt2011} and the {AMMeter}\cite{makonin2013} stressing several aspects similar to {\ac{YoMo}}.
In Table \ref{tab:openSourceMeter} a review on open-source and research smart meters are presented and compared to the proposed {\ac{YoMo}} metering approach.
Distinctive features are the used communication methods, the measured quantities, the number of measurable devices/connections, the switch-ability, the sampling frequency of the measurements, the way the power calculations are performed, the open-source and open-hardware availability, if the board is Arduino-based and the approximated costs of the metering unit.

\begin{landscape} 
 \begin{table*}
 \centering
 \begin{tabular}{c|ccccccc}
\hline
 & CSM &OEM & AMMeter  & ACme-A  & ACme-a & SmartMeter.Kom &  \ac{YoMo}\\
\hline
Communication & Ethernet	& SUB 1GHz & x & ZigBee & ZigBee & ZigBee &Wi-Fi\\
Measurement	& P,Q,S,I,V,E &  P/S & I & P/Q/S & S & I,V &P,Q,S,I,V,E\\
Number of Connections & 1 & 3 & 12 & 1 & 1 & 1 & 1\\
Intrusiveness	& - & $\surd$ & $\surd$ & x & x & - & x\\
Switchable	& x &x & x & $\surd$ & $\surd$ & $\surd$ & $\surd$\\
Sampling Frequency & - &  variable & 1Hz & 1Hz & variable & - & variable   \\
Power calculation	& hardware & software & software & hardware & software & - & hardware\\
Open-source & x &$\surd$ & $\surd$ & $\surd$ & $\surd$  & x &$\surd$ \\
Arduino-based & $\surd$ & $\surd$ & $\surd$ & x & x & x & $\surd$ \\
Approximated costs & \EUR{100}  & \EUR{55} & x  & x & x  & x & \EUR{65}\\
\hline
\end{tabular}
\caption{Comparison of existing open-source and research metering solutions}
\label{tab:openSourceMeter}
\end{table*}
\end{landscape}

\section{Smart Metering Concepts at a Glance} \label{sec:measurement}
A smart meter consists of different units such as the used metering approach, how safety of the meter is achieved, how attached connections can be switched and how data between the smart meter and user can be communicated.
These main components with its variations and possibilities are presented and discussed in the following section.
\subsection{Voltage and Current Sensing} \label{sec:sensing}
To calculate reactive and active power for a connected load, three physical quantities are required: the voltage, u, the current, i, and the phase shift between them, $\varphi_{ui}$. The voltage can be easily determined using a voltage transformer or a voltage divider feed into a ADC\footnote{ADC - analog digital converter} input or energy monitoring IC.
Numerous current measurement methods exist but with most of them a compromise has to be made between resolution and accuracy.
These can be divided into two kinds where the current either is transformed into a useful scale or another physical quantity.
A current along a wire invokes a voltage drop and a magnetic field around/inside the wire.
A possible sensing technique is the current shunt resistor.
The voltage drop across the shunt is used as a proportional measure of the current flow.
Resistors incur a power loss proportional to the square of the current \cite{Ziegler:2009}.
The best proved and most applied methods build on the magnetic field principle \cite{kirkham:2009}.
An advantage of this method is the galvanic isolation between input and output of the current transformer, allowing the circuit connected to the output to be made touch safe:
One possibility are Magnetoresistant (MR) materials changing their electrical resistance when a magnetic field is applied to them.
This effect is not linear and is temperature dependent, which makes it complicated to use for a wide current range.
An additional error source is remanent magnetization, which limits the dynamic DC range.
  Using a bridge configuration consisting of multiple sensors, it is possible to compensate for this unwanted behaviour.
  This method is used to measure currents up to 50A. Typical measurement errors with this method are 0.3\% \cite{ripka:1996}.
A further possibility is the Hall effect describing the production of a voltage signal across an electrical conductor.
The current signal is applied to a magnetic core.
The resulting core flux is sensed by the Hall element, which is placed in the magnetic core.
This setup is very similar to the current transformer, with the difference being that the flux sensing is done by a transformer winding instead of the Hall element.
Remanent magnetization in the sensor is a source of error here.
The final possibility is the Current transformer method exploiting Faraday`s law of induction. The current signal is converted to a useful scale by a transformer with an appropriate gear ratio. Further processing is done by ICs or logic.
The phase shift, $\phi_{ui}$, is defined as $\phi_{ui} = \phi_{u} - \phi_{i}$ and can also be expressed as the time shift between voltage and current. Sensing circuits such as phase-locked loops \cite{PLL:2012} or numerical methods such as the cross-correlation are used to estimate this quantity.

\subsection{Power Calculation} \label{sec:powercalc}
A typical feature of a smart metering system is to determine the power at a load and monitor the energy consumption over time.
In Section~\ref{sec:sensing} we discussed in detail how to obtain information about current, voltage, and the time shift between them. From these parameters we can derive the following physical quantities:
\begin{itemize}
\item Active power: $P_W = U_{RMS} \cdot I_{RMS} \cdot cos(\phi_{ui})$
\item Reactive power: $P_B = U_{RMS} \cdot I_{RMS} \cdot sin(\phi_{ui})$
\item Apparent power, $P_S = U_{RMS} \cdot I_{RMS}$
\end{itemize}
These definitions show that the root mean square (RMS) values for current and voltage as well as the phase shift between them have to be determined. The simplest way to implement RMS sensing is done by multiplying the peak amplitude with the crest factor, defined by $i_{RMS} = \sqrtsign{2} \cdot \hat{i}$ for a sinusoidal signal. This method is imprecise when the signal deviates from a steady-state sinusoidal signal.\\
Monitoring energy consumption over a given interval $T=[t_0,t_1]$ with step size $T_s = \frac{1}{f_s}$ can be implemented by approximating the integration of power over time.
\begin{equation*}
E = \int_{t_0}^{t_1} P(t) dt \sim \sum_{i=t_0}^{t_1} P(i) \cdot T_s
\end{equation*}
The resolution is highly dependent on the sampling frequency, $f_s$, which has to be well-chosen because every load's time behaviour differs. Therefore, the sensing logic should dynamically adapt the sampling frequency due to a connected load. For a near-constant load, the sampling frequency could be set very low, but not less than $100Hz$ due to the Nyquist-Shannon sampling theorem.
A smaller sampling frequency results in a smart meter with a smaller energy consumption because in this inoperational state, a further step would include setting the whole sensing logic to sleep mode in order to save energy.

\subsection{Galvanic Isolation} \label{sec:galvanic}
An important aspect in grid metering is isolation between the grids ground potential and the ground potential of the metering circuit. The occurrence of high voltage peaks or ground loops may destroy the whole setup.  Integrating transformers or transducers into the measurement circuit is a popular and well-proven way to achieve this isolation. The optical coupler is used for galvanic isolation between integrated circuits. Consisting of a light-emitting diode at the input and a phototransistor at the output, it uses light to transmit electrical signals and has a very small form factor.

In terms of current measurement some setups such as the current transformer, Hall effect sensors, or MR sensors already provide galvanic isolation. In contrast, the shunt method does have the grid's ground electrically connected to the measurement unit, which makes the approach less safe. Voltage sensing setups are usually implemented using a voltage divider. A transducer between voltage divider and grid potential seems to be the best way to achieve galvanic isolation.

\subsection{Load Switching} \label{sec:switching}
The ability to switch a connected load on and off has become an important requirement for smart meters. Loads can be switched remotely by the smart meter, which allows scheduling jobs at times when the energy price is lower. The runtime of high-power appliances is moved to periods with the lowest energy price to help consumers reduce their electricity bill.
Load switching is achieved by inserting solid state relays or conventional relays into the current path of the load. These electrical parts have special inputs that can be driven by a microcontroller or any other smart device.

A conventional CEE 7/4 Schuko plug socket does not have a well-defined line and neutral layout. Therefore, a potential load switching implementation must switch both circuits: line and neutral wire.

\subsection{Communication Unit}
An important aspect of a state-of-the-art smart meter is its capability to communicate.
To ensure this need, modern smart meters offer a wired (e.g., power line communication, fieldbuses) or wireless (ZigBee, WLAN, cellular communication) communication possibilities.
Possible technologies mainly differ on its data rates, transmission range, frequency bands and number of plugged nodes.
In the context of homes, ZigBee or WLAN are preferable technologies because of their low cost and simplicity for integration.

\section{\ac{YoMo} Smart Meter}
According to Section \ref{sec:measurement} presenting the main parts of a smart meter, we present in the following our design choices for the proposed metering approach.
\begin{itemize}
\item \textbf{Relay}: An essential requirement for a smart metering system has become the ability to switch the load's state on and off. The upcoming question here is either to assemble a solid state relay or a mechanical one. Indeed heating up is an influential factor for this decision. The current flow in the device leads to dissipation power. In a solid state relay this power is about $2.5W$, which leads to the device heating up and hence requires a heat sink. This heat sink, of course, needs also to be mounted on the shield and requires additional space. The forward current of a conventional solid state relay is about $20mA$\footnote{Sharp S216S02 Series} and therefore about five times smaller than the forward current of a mechanical relay. The big advantage of a mechanical relay is the ability to switch currents up to 20A, which covers all single-phase household devices. For this reason we integrated a mechanical relay in our design. This relay is driven by one of the Arduinos analog outputs. Unfortunately the forward current is about $100mA$, so the relay consumes a half Watt. The resulting advantage is that we are able to switch phase and neutral conductor at the same time. Both, mechanical and solid state relay, can be controlled by the Arduino's analog output. This output integrated in the firmware enables the user to control the connected load via the Raspberry Pi's homepage.
\item \textbf{Sensors}: For estimating the energy consumption two electrical quantities have to be determined: voltage and current. Voltage measurement is easily implemented by a voltage divider. The big problem here is that the potential forwarded to the energy monitor is not galvanically isolated from the grids potential. An efficient solution provides the isolation amplifier. Connected to the smaller dividers resistor he outputs a galvanically isolated signal that is in phase with the input. The differential output of the isolation amplifier is connected to the energy monitors input channel.\\Numerous sensors for estimating the current flow exist. To meet the challenge metering currents from $mA$ range up to 20A only a few technologies are utilisable. Our metering shield is equipped with a current transformer. A current transformer outputs a galvanically isolated signal and therefore suits our concept. This signal is scaled down and forwarded to the energy monitor.
\item \textbf{Energy monitor}: An energy monitor is an integrated circuit responsible for calculating physical quantities such as active power, reactive power and apparent power from given input signals. Our metering shield contains the ADE7753 chip. This chip is connected to the isolation amplifier, current transformer, and the Arduino UNO board. This IC calculates all necessary electrical quantities such as the consumed energy of the connected load from the signals provided by the connected sensors. As necessary electrical quantities we defined: $V_{RMS}$, $I_{RMS}$, reactive power, active power, and apparent power. These quantities are sampled due to the given sampling frequency that is provided to the system by the user. The estimated values are written to registers inside the ADE7753. These registers are read out by the Arduino board via serial communication such as SPI.
\item \textbf{Arduino UNO board}: The used Arduino board is not part of our metering shield, but enumerating it here helps to understand the interaction of the systems components. The metering component is indeed the metering shield, which is connected to the Arduino via serial communication and other control signals. Our design concept does not depend on a certain Arduino board. Since the Arduino UNO has the smallest form factor and the lowest price this choice is reasonable. The Arduino's first task is to read out register values of the energy monitor to a given sample interval and forward the estimated data to the coordination device. The second task is to parse incoming commands from the coordination device and to run them. The implemented command set includes commands to switch the connected load, set the energy monitor to sleep and wake it, and to adjust the sampling frequency.
\end{itemize}
The \ac{YoMo} board described as a system overview, unsoldered PCB board and equipped PCB Board are presented in Figure \ref{fig:shSchem}.
\section{\ac{YoMo} enabled Smart Meter System}

Beside the introduction of the \ac{YoMo} metering shield,  we aimed at a system that combines low-cost hardware that fulfills requirements such as real-time metering, providing data about active, reactive, apparent power, voltage and current, switching connected circuits on and off, adjusting of the used sampling frequency and the ability to display measured data to the user.
\begin{figure}[h]
 \centerline{\includegraphics[width=1\columnwidth]{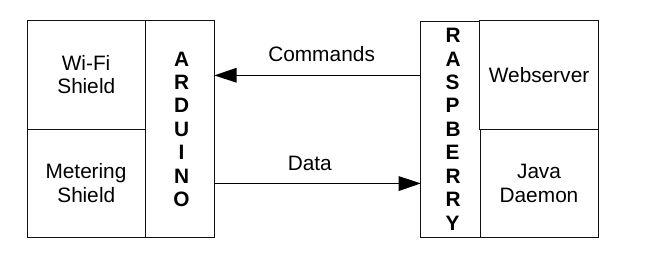}}
  \caption{System overview: Arduino board and Raspberry Pi form the system. Arduino measures and forwards energy data. Raspberry Pi  and receives commands from Raspberry Pi.}
  \label{fig:sysSchem}
\end{figure}
Therefore we chose the Arduino UNO as metering device and the Raspberry Pi as device to supervise the metering process and display the estimated measurement data. Figure \ref{fig:sysSchem} summarizes our concept.
\begin{figure*}
\subfigure[Measurement chain: Relay, sensors and, energy monitor are located on the metering shield]
 {\includegraphics[width=0.8\columnwidth]{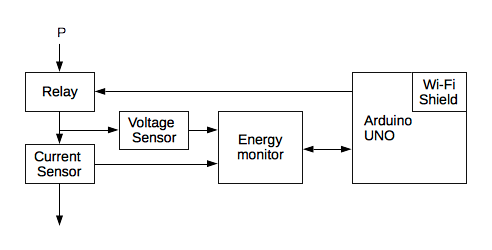}}
\hfill
\subfigure[\ac{YoMo} metering shield not soldered with electronic parts]{\includegraphics[width=0.6\columnwidth]{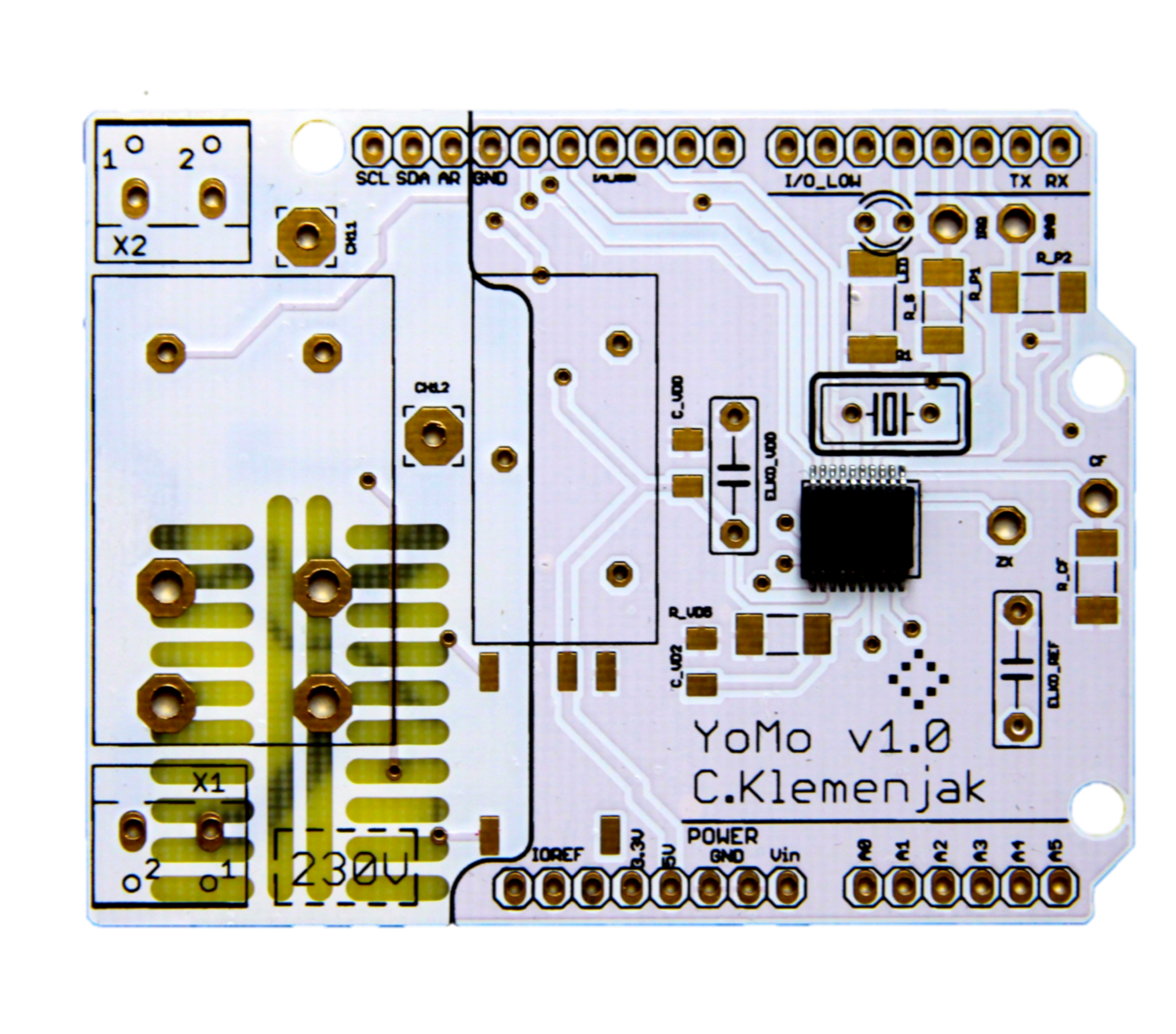}}
\hfill
\subfigure[\ac{YoMo} metering shield soldered with electronic parts]{\includegraphics[width=0.6\columnwidth]{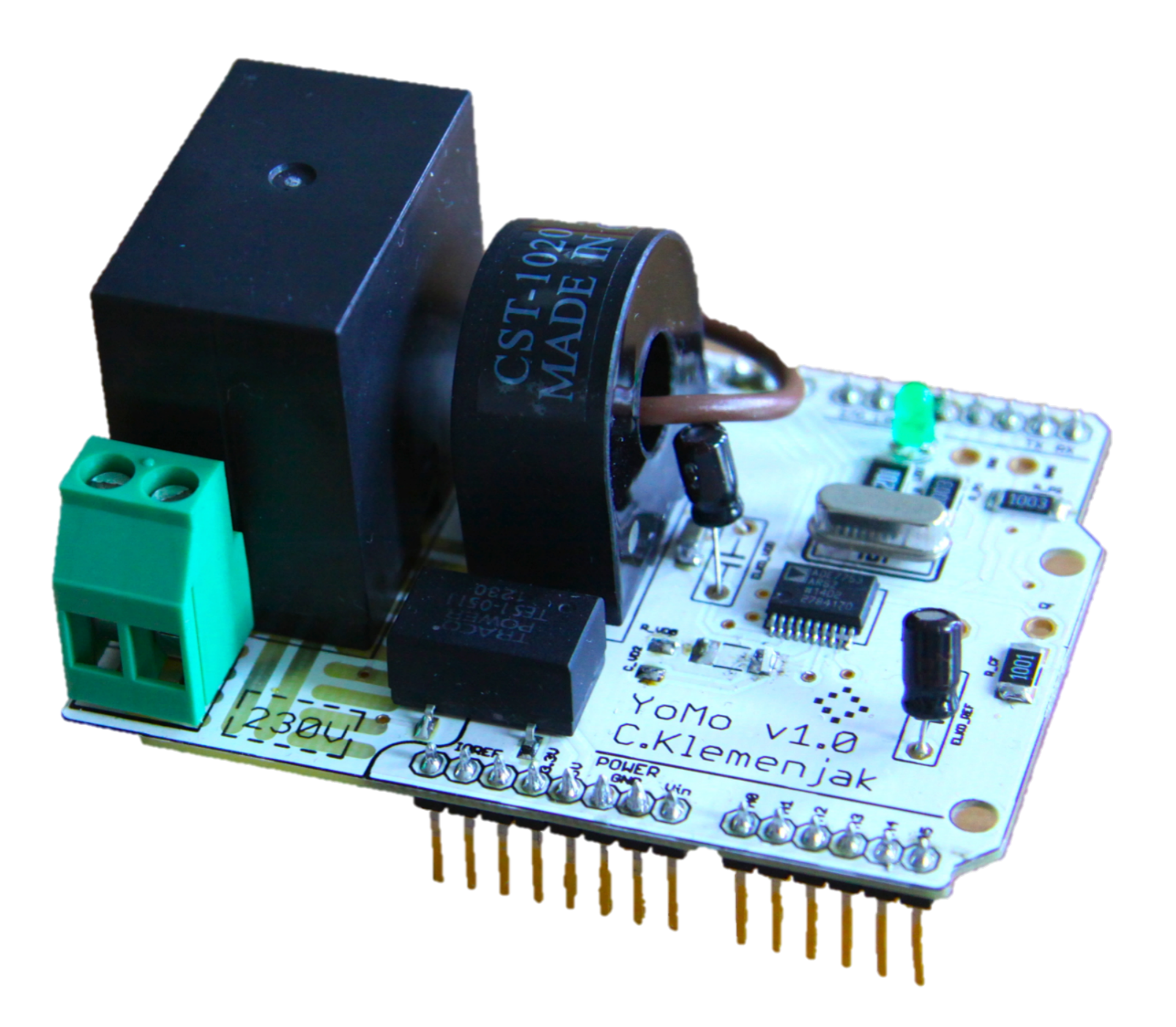}}

\hfill
\caption{The \ac{YoMo} metering shield}
\label{fig:shSchem}
\end{figure*}
This concept can be split up into 2 parts:
 \begin{itemize}
\item \textbf{Metering device (Arduino\footnote{http://arduino.cc/})}: The first device described is the smart meter itself, in our case the Arduino UNO board with its extension shields (Wi-Fi and Metering named \ac{YoMo}).
Electrical quantities such as the RMS values of current and voltage as well as the power quantities are estimated to a given sampling frequency.
Since the Arduino is equipped with an Wi-Fi shield, we use the NTP\footnote{NTP - network time protocol} service to timestamp our data.
This data is stored local on a SD card first and second being transmitted via Wi-Fi to the coordination device.
The program schedule contains a well-defined time window where the metering device checks if commands where transmitted from the coordination device.
Depending on the command the metering device switches the connected load on or off, enables/disables the energy monitor on the metering shield or adjusts the sampling frequency.
\item \textbf{Coordination device (Raspberry Pi\footnote{http://www.raspberrypi.org/})}: The second device described is the coordination device, in our case the Raspberry Pi.
We developed a Java daemon that listens on a UDP\footnote{UDP - user datagram protocol} port for incoming measurement data transmitted by the metering device.
These data is read, checked and saved with a unique identifier.
The coordination device runs a web server, on which the measurement data is accessible for every host, meaning personal computers, smartphones and tablets via a simple graphical user interface.
The index page contains JavaScript elements which plot data about energy consumption over time.
Also embedded are control elements and an input field for the sampling frequency.
These commands are caught by the Java daemon and sent to the metering device.
This way it is possible to control the metering system as well as the attached load(s) via a conventional web browser.
The metering system supports multiple metering devices, so that a complete smart metering system may be formed with a central coordinator that gathers all data about energy consumption.
\end{itemize}

The centrepiece of a smart meter is the sensing unit that measures the desired physical quantities.
The implementation of this special unit varies from meter to meter.
The design is highly depending on the set requirements.
Our requirements were to achieve galvanic isolation, the ability to meter currents up to $20A$, to be switchable, to adjust the sampling frequency as well as to build an energy monitor that estimates all energy quantities.
With these design goals in mind we designed a Arduino-compatible metering shield, which is summarized in Figure \ref{fig:shSchem}.
This shield is independent on the used Arduino board and provides access to all estimated physical quantities via the {SPI}\footnote{SPI - serial peripheral interface} bus as well as access to the built-in relay.

\section{Case Studies}

\begin{table}[h]
  \begin{center}
  \begin{tabular}{lcccc}
    \hline
    Household appliance & Measured value & Real value    &  Error \\ \hline 
    
refrigerator & 50  & 52  & 3.8 \%\\
ventilator & 83  &  81  & 2.47 \%\\
convection oven & 729  & 752  & 3.05 \%\\
water kettle & 1910  & 1930  & 1.03 \%\\ 
radiant heater & 1989  & 1980  & 0.51 \%\\ \hline

  \end{tabular}
\caption{Active power measurement in [W]}
\label{powermeteringsetup}
\end{center}
\end{table}

\begin{table}[h]
  \begin{center}
  \begin{tabular}{lcccc}
    \hline
    Household appliance & Measured value & Real value & Error \\ \hline 
    refrigerator & 95 & 99 & 4.04 \%\\
    ventilator & 85 & 88 & 3.4 \%\\
    convection oven & 734 & 753 & 2.5 \%\\
       water kettle & 1936 & 1940 & 0.2 \%\\ 
    radiant heater & 2049 & 2000 & 2.45 \%\\
 \hline
  \end{tabular}
\caption{Apparent power measurement in [VA]}
\label{apparentpowermeteringsetup}
\end{center}
\end{table}

\begin{table}[h]
  \begin{center}
  \begin{tabular}{lcccc}
    \hline
    Household appliance & Measured value & Real value    &  Error \\ \hline 
    ventilator & 39  & 36 & 8.3 \%\\
    refrigerator & 85 & 84 & 1.2 \%\\
  \end{tabular}
\caption{Reactive power measurement in [var] }
\label{reactivepowermeteringsetup}
\end{center}
\end{table}

\subsection{Measurement Accuracy}
To evaluate the accuracy of the \ac{YoMo} metering board, we measured the active, the reactive\footnote{Only presented for appliances having a reactive power quantity} and the apparent power of different typical household appliances and compared the results with measurements by the power measurement unit of an isolating transformer (Block {brs2200}) which has an measurement accuracy of $+/-3\%$.
Table \ref{powermeteringsetup}, \ref{apparentpowermeteringsetup} and \ref{reactivepowermeteringsetup}	present the measured values by \ac{YoMo} against the reference measurement unit and provides the measurement error for each device.
According to the results, we claim that \ac{YoMo} is providing sufficient accuracy to monitor household appliances of a nominal power from a few watts up to some kilowatts.

\subsection{Energy Monitoring System based on \ac{YoMo} and Raspberry Pi}
\begin{figure}[htpb]
 \includegraphics[width=1\columnwidth]{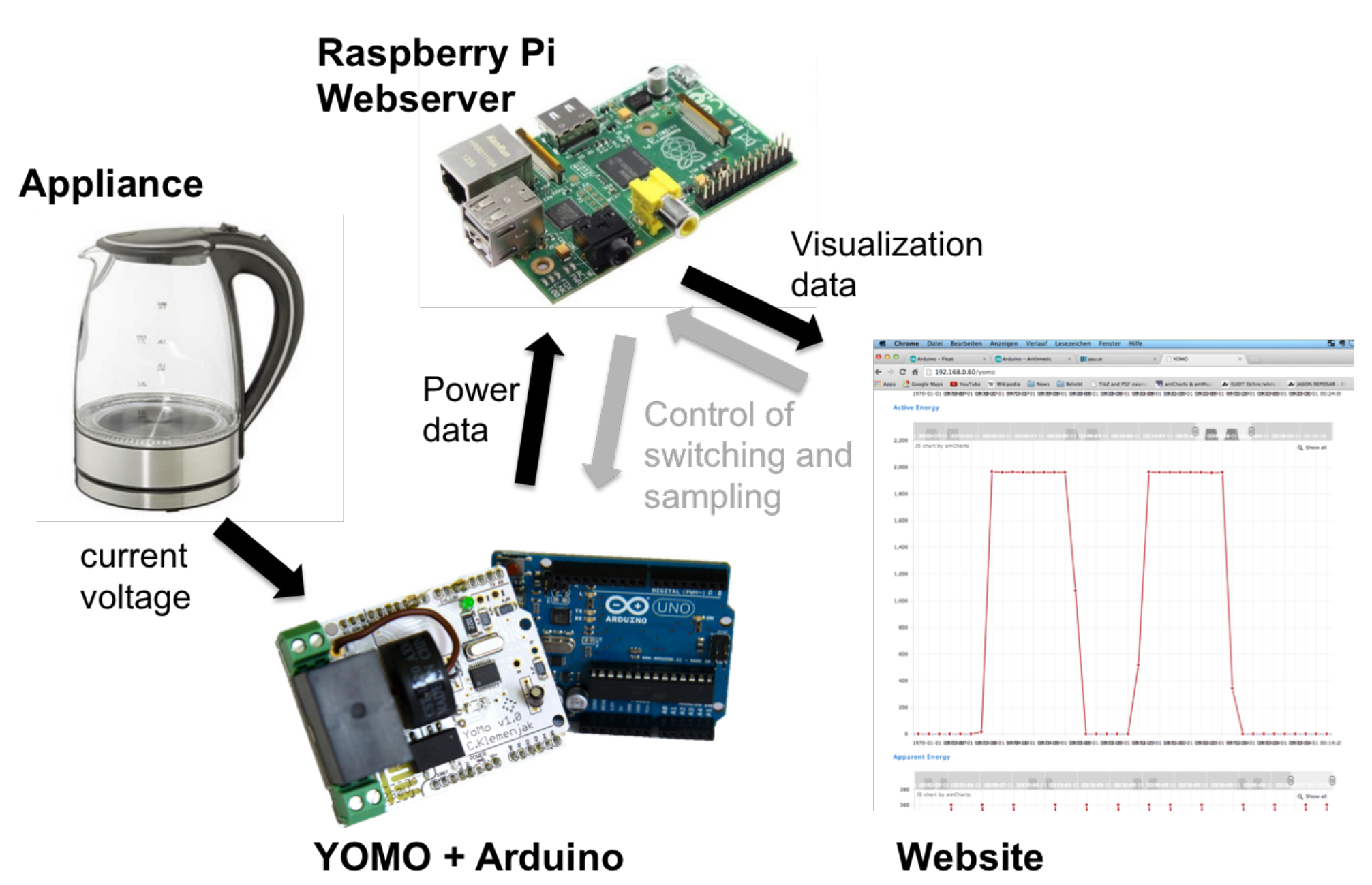}
  \caption{System overview of a simple home energy management system with the Arduino board, the {YoMo} metering shield and Raspberry Pi}
  \label{fig:YoMoPi}
\end{figure}

The general system overview of the Arduino board extended with {\ac{YoMo}} and with the Raspberry Pi acting as visualization and controlling unit is shown in Figure \ref{fig:YoMoPi}.
It can be seen that on the Raspberry Pi a webserver is running which is able to visualize current and historical measurement data.
Via the website it is also possible to communicate with {\ac{YoMo}} to either control the connected device or to adjust the used sampling frequency.
The sketched power draw is from a water kettle with a stated power demand of 2000W.

\section{Future Work}
One major topic is to tune the accuracy of the metering board by evaluating its performance with more accurate reference measurement units and controllable loads.
To improve the proposed open-source smart metering approach one major topic is it how to establish the connection to the quantity to be measured.
It is possible to measure either an appliance or a group of appliances by a smart plug connection or to measure the whole household grid or circuits of the grid by current terminals.
Respectively, another aim for improving the meter is extending the board from a single-phase metering approach to a three-phase metering approach.
For example, Austria, uses a three-phase system in which the major appliance are driven with one phase, but appliances such as a stove can be operated by three phases \cite{Monacchi2013}.
Finally, by adjusting the design in a way that {\ac{YoMo}} can operate via standard interfaces such as {SPI} with other development boards such as the {Raspberry Pi} or {BeagleBoard}\footnote{http://beagleboard.org/} would improve the applicability of board.
\ac{YoMo} could then be used as an extension board for application more computational demanding such as applications for  Non-Intrusive Load Monitoring \cite{Egarter2014} and demand response issues.

\section{Conclusion\label{sec:conclusion}}
In this paper a novel open source and open hardware metering approach has been presented and discussed.
We present how the proposed metering approach differs from other existing open-source and research metering approaches by distinctive features such as the measured quantities, the number of measurable connections/devices, the ability to switch loads, the used sampling frequency, etc.
The introduced metering board, named as {\ac{YoMo}}, is designed as an Arduino Shield which is able to measure electric quantities such as active power, reactive power, apparent power, current and voltage and can switch appliances up to 20A.
We demonstrate a metering system including a Raspberry Pi acting as a coordination unit and web server to visualize the measured quantities and to provide the possibility to switch the connected circuit and to adjust the used sampling frequency.
The {\ac{YoMo}} metering approach was designed as open-source and open-hardware with the aim to design a low-cost smart meter which is both easy to use and easy to extend and modify according to the current application.

\section{Acknowledgments}
This work was supported by Lakeside Labs GmbH, Klagenfurt, Austria and funding from the European Regional Development Fund and the Carinthian Economic Promotion Fund (KWF) under grant KWF-$20214\mid22935\mid24445$.

\bibliographystyle{IEEEtran}
\bibliography{bibfile}

\end{document}